\documentclass[preprint,aps,eqsecnum]{revtex4}
\usepackage{graphicx}
\newcommand{\be}{\begin{eqnarray}}
\newcommand{\ee}{\end{eqnarray}}

\begin{document}
\title
{Chiral Odd Generalized Parton Distributions in Impact Parameter Space}

\author{\bf Harleen  Dahiya$^a$, Asmita Mukherjee$^b$}
\affiliation{$^a$ Department of Physics, National Institute of
Technology, Jalandhar, Punjab 144011, India\\ $^b$ Department of
Physics, Indian Institute of Technology, Powai, Mumbai 400076,
India.}
\date{\today}
\begin{abstract}
We investigate the chiral odd generalized parton distributions (GPDs)
for the quantum fluctuations of an
electron in QED. This provides a field theory inspired model of a
relativistic spin $1/2$ composite state with the correct correlation
between the different light-front wave functions (LFWFs) in Fock space. 
We express the GPDs in terms of overlaps of LFWFs and obtain their
representation in impact parameter space when the momentum transfer is purely
transverse. We show the spin-orbit correlation effect of the two-particle
LFWF as well as the correlation between the constituent spin and the
transverse spin of the target.
\end{abstract}
\maketitle
\section{Introduction}
Generalized parton distributions (GPDs) contain a wealth of
information about the nucleon structure (see \cite{rev} for
example). At zero skewness $\xi$, if one performs a Fourier
transform (FT) of the GPDs with respect to (wrt) the momentum
transfer in the transverse direction $\Delta_\perp$, one gets the
so called impact parameter dependent parton distributions
(ipdpdfs), which give how the partons of a given longitudinal
momentum are distributed in transverse position (or impact
parameter $b_\perp$) space. These obey certain positivity
constraints and unlike the GPDs themselves, have probabilistic
interpretation \cite{bur}. The $x$ moment of the GPDs give the
nucleon form factors. The ipdpdfs were first introduced in the
context of the form factors in \cite{soper}. Ipdpdfs are defined
for nucleon states localized in the transverse position space at
$R_\perp$. In order to avoid a singular normalization constant,
one can take a wave packet state. A wave packet state which is
transversely polarized is shifted sideways in the impact parameter
space \cite{burchi}. This gives an interesting interpretation of
Ji's angular momentum sum rule \cite {ji}: the expectation value
of the transverse spin operator receives contribution from the
second $x$ moment of both the GPDs $H(x,0,0)$ as well as
$E(x,0,0)$; the term containing $E(x,0,0)$ arises due to a
transverse deformation of the GPDs in the center of momentum frame
and the term containing $H(x,0,0)$ arises due to an overall
transverse shift when going from transversely polarized nucleons
in the instant form to the front form.

At leading twist, there are three forward parton distributions
(pdfs), namely, the unpolarized, helicity and transversity
distribution. Similarly, three leading twist generalized quark
distributions can be defined which in the forward limit, reduce to
these three forward pdfs. The third one is chiral odd and is
called the generalized transversity distribution $F_T$. It is
parametrized in terms of four GPDs, namely $H_T$, $\tilde H_T$,
$E_T$ and $\tilde E_T$ in  the most general way \cite{markus,chiral,burchi}. 
Unlike $E$, which
gives a sideways shift in the unpolarized quark density in a
transversely polarized nucleon, the chiral-odd GPDS affect the
transversely polarized quark distribution both in unpolarized and
in transversely polarized nucleon in various ways. $\tilde E_T$
does not contribute when skewness $\xi=0$, as it is an odd
function of $\xi$. $H_T$ reduces to the transversity distribution
in the forward limit when the momentum transfer is zero. Unlike
the chiral even GPDs, information about which can be and has been
obtained from deeply virtual Compton scattering and hard exclusive
meson production, it is very difficult to measure the chiral odd
GPDs. At present there is only one proposal to get access to them
through diffractive double meson production \cite{double}. There
is also  a prospect of gaining information about their Mellin
moments from lattice QCD. They have been investigated in a constituent
quark model in \cite{barbara}, where a model independent overlap in terms of
LFWFs is also given. However, the chiral odd GPDs provide
valuable information on the correlation between the spin and
angular momentum of quarks inside the proton \cite{chiral} and so
it is worthwhile to investigate their general properties. 

The impact representation of GPDs has been extended to the chirally odd sector
in \cite{chiral}. In this work, we investigate the chiral odd GPDs for
the quantum fluctuations  of a  lepton in QED at
one-loop order \cite{drell}, the same system which gives the Schwinger
anomalous moment $\alpha/2 \pi$. One can generalize this analysis
by assigning a mass $M$ to the external electrons and a different
mass $m$ to the internal electron lines and a mass $\lambda$ to the
internal photon lines with $M < m + \lambda$ for stability. In this work, we
use $M=m$ and $\lambda=0$. In effect, we
shall represent a
spin-${1\over 2}$ system as a composite of a spin-${1\over 2}$
fermion and a spin-$1$ vector boson \cite{dis,dip,dip2,marc,si}. 
This model has the
advantage that it is lorentz invariant, and has the correct correlation
between the Fock components of the state as governed by the light-front
eigenvalue equation. Also, it gives an intuitive
understanding of the spin and orbital angular momentum of a
composite relativistic system \cite{orbit}. In the light-front gauge
$A^+=0$, the GPDs are
expressed as overlaps of the light-front
wave functions (LFWFs).  Because of Lorentz invariance, $\xi$ dependence
of the $x$ moment of the GPDs gets canceled between the $2-2$ and $3-1$
overlaps and one automatically gets the form factors as a function of
the momentum transfer squared \cite{overlap}.  We take the skewness to be
zero.

The plan of the paper is as follows. In section II we calculate the chiral
odd GPDs for a dressed electron state at one loop in QED. In section III we
calculate the corresponding ipdpdfs. Conclusions are presented in section
IV.
\section{Chiral odd generalized parton distributions}
The chiral odd generalized quark distribution is parametrized as :
\be
F^{j}_T(x,\xi,t) &=& -i \int \frac{dy^-}{8\pi}\, e^{i x P^+
\frac{y^-}{2}}\, \langle P',\sigma'|\, \bar{\psi}(-\frac{y^-}{2})
 \sigma^{+j} \gamma_5\psi(\frac{y^-}{2}) |\,P,\sigma\rangle \,
\nonumber \\
 &=& -\frac{i}{2P^+} \left[ H_T(x,\xi,t)\, \bar{u}
\sigma^{+j} \gamma_5\, u
  + \tilde{H}_T(x,\xi,t)\, \bar{u} \frac{\epsilon^{+j\alpha\beta}
    \Delta_\alpha P_\beta}{m^2} u \right .
\nonumber \\ & & \left . + E_T(x,\xi,t)\, \bar{u}
\frac{\epsilon^{+j\alpha\beta}
    \Delta_\alpha \gamma_\beta}{2m} u
  + \tilde{E}_T(x,\xi,t)\, \bar{u} \frac{\epsilon^{+j\alpha\beta}
    P_\alpha \gamma_\beta}{m} u
  \, \right].
\label{transF}
\end{eqnarray}
We have omitted the helicity indices of the spinors.
Here $j=1,2$ are the transverse components, $u(P)$ and $\bar{u} (P')$ are
the initial and final state proton spinors respectively. The totally
antisymmetric tensor is given by $\epsilon^{+-12}=-2$. We use kinematical
variables ${\bar P}=(1/2) (P+P')$, $\Delta=P'-P$,
$t=\Delta^2=-(\Delta_\perp)^2$. The {\it r.h.s } can be calculated
using the LF spinors
\be
u_{\uparrow}(p)={1 \over \sqrt{2 p^{+}}}
\left(\begin{array}{c}p^{+} + m \\ p^{1}+ip^{2}\\
p^{+}-m \\ p^{1}+ip^{2} \end{array}\right)\,,
\ee

\begin{eqnarray}
u_{\downarrow}(p)={1 \over \sqrt{2 p^{+}}}
\left(\begin{array}{c}-p^{1} + i p^{2}\\p^{+} + m \\ p^{1}-ip^{2}\\
-p^{+}+m  \end{array}\right)\,,
\end{eqnarray}
where $m$ is the mass of the fermion.  In order to calculate the {\it l. h. s.}
of the Eq. (\ref{transF}), we use the two-component formalism in
\cite{two}. The `good' light-front (LF) components of the fermion
field are projected by $\psi^{\pm}=\Lambda^{\pm}\psi$ with
$\Lambda^{\pm}={\frac{1}{2}} \gamma^{0}\gamma^{\pm}$;
$\gamma^{\pm}=\gamma^0 \pm \gamma^3$. By taking the appropriate
$\gamma$-matrix representation, one can write
\be
    \psi^+(y)= \left[\begin{array}{cc} \zeta(y) \\ 0
\end{array}
        \right]\,
\ee
with $\zeta$ being a two-component field.

We take the state $ \mid P, \sigma \rangle$
of momentum $P$ and helicity $\sigma$ to be a dressed electron
consisting of bare states of an electron and an electron plus
a photon :
\begin{eqnarray}
\mid P, \sigma \rangle && = {\cal N}\Big [ b^\dagger(P,\sigma) \mid 0
\rangle
\nonumber \\
&& + \sum_{\sigma_1,\lambda_2} \int
{dk_1^+ d^2k_1^\perp \over \sqrt{2 (2 \pi)^3 k_1^+}}
\int
{dk_2^+ d^2k_2^\perp \over \sqrt{2 (2 \pi)^3 k_2^+}}
\sqrt{2 (2 \pi)^3 P^+} \delta^3(P-k_1-k_2) \nonumber \\
&& ~~~~~\phi_2(P,\sigma \mid k_1, \sigma_1; k_2 , \lambda_2) b^\dagger(k_1,
\sigma_1) a^\dagger(k_2, \lambda_2) \mid 0 \rangle \Big ].
\label{eq2}
\end{eqnarray}

Here $a^\dagger$ and $b^\dagger$ are bare photon and electron
creation operators respectively and $\phi_2$ is the
two-parton wave function. It is the probability amplitude to find
one electron plus photon inside the dressed electron state.

We introduce Jacobi momenta $x_i$, ${q_i}^\perp$ such that $\sum_i x_i=1$
and
$\sum_i {q_i}^\perp=0$.  They are defined as
\be
x_i={k_i^+\over P^+}, ~~~~~~q_i^\perp=k_i^\perp-x_i P^\perp.
\ee
Also, we introduce the wave function,
\be
\psi_2(x_i,q_i^\perp)= {\sqrt {P^+}} \phi_2 (k_i^+,{k_i}^\perp);
\ee
which is independent of the total transverse momentum $P^\perp$ of the
state and  boost invariant.
The state is normalized as,
\be
\langle P',\lambda'\mid P,\lambda \rangle = 2(2\pi)^3
P^+\delta_{\lambda,\lambda'} \delta(P^+-{P'}^+)\delta^2(P^\perp-P'^\perp).
\label{norm}
\ee
The two particle wave function depends on the helicities of the electron and
photon. Using the eigenvalue equation for the light-cone Hamiltonian, this
can be written as \cite{dis},
\be
\psi^\sigma_{2\sigma_1,\lambda}(x,q^\perp)&=& -{x(1-x)\over
(q^\perp)^2+m^2 (1-x)^2} {1\over {\sqrt {(1-x)}}} {e\over {\sqrt
{2(2\pi)^3}}} \chi^\dagger_{\sigma_1}\Big [ 2 {q^\perp\over
{1-x}}+{{\tilde \sigma^\perp}\cdot q^\perp\over x} {\tilde
\sigma^\perp} \nonumber\\&&~~~~~~~~~~~~~~~~~~ -i m{\tilde
\sigma}^\perp {(1-x)\over x}\Big ]\chi_\sigma \epsilon^{\perp
*}_\lambda {\cal N}. 
\label{psi2} \ee $m$ is the bare mass of the
electron, $\tilde \sigma^2 = -\sigma^1$ and $\tilde \sigma^1=
\sigma^2$. ${\cal N}$  gives the normalization of the state.
 $\chi_{\sigma}$ is the two component
spinor for the electron and
$\epsilon_{\lambda}^{\perp}$ is the polarization vector of the photon.

For $\xi=0$, the momentum transfer is purely transverse,
\be
t=(P-P')^2=-\Delta_\perp^2\,. \ee

The two-particle contribution to the off forward matrix element is given in
terms of overlaps of $\psi^{*\sigma'}_{2\sigma'_1,\lambda'}(x',k'^\perp)$ and
$\psi^{\sigma}_{2\sigma_1,\lambda}(x,k^\perp)$,
where
\begin{equation}
{k'_\perp}={k_\perp}-(1-x)\ {\Delta_\perp}~~~~~{\rm and}~~~~
x'=x\,;
\end{equation}
where $a_\perp=-a^\perp$.
As $\xi=0$, there are no particle number changing overlaps. 

Eq. (\ref{eq2}) represents a state having definite momentum and light-front
helicity. The transversely polarized states can be expressed in terms of 
the helicity states as
\be
\mid x \rangle = {1\over \sqrt{2}} ( \mid \uparrow \rangle + \mid
\downarrow \rangle ) ; ~~~~~~\mid y \rangle = {1\over \sqrt{2}} (
\mid \uparrow \rangle + i \mid \downarrow \rangle ) ;\ee where
$\mid x \rangle$ and  $\mid y \rangle$ denote states polarized in
the $x$ and $y$ directions, respectively and $\mid \uparrow
(\downarrow) \rangle$ denotes states with positive (negative)
helicity. The overlaps can be calculated for different helicity
configurations using the two-particle wave function given above.
We get,
\begin{eqnarray}
F^{1}_T(\uparrow\downarrow) & = &
H_T(x,0,t)+\frac{\tilde{H}_T(x,0,t)}{2 m^2}
(i\Delta_2)(\Delta_1-i\Delta_2) \nonumber \\
&=& \frac{e^2}{(2\pi)^3}\frac{x}{1-x}\left[I_1+I_2+C I_3\right]\,,
\\
F^{1}_T(\downarrow\uparrow) & = &  H_T(x,0,t)
+\frac{\tilde{H}_T(x,0,t)}{2 m^2} (-i\Delta_2)(\Delta_1+i\Delta_2)
\nonumber \\
&=& \frac{e^2}{(2\pi)^3}\frac{x}{1-x}\left[I_1+I_2+C I_3\right]\,,
\\
F^{1}_T(\uparrow\uparrow) & = & \frac{1}{2m}[E_T(x,0,t)+2
\tilde{H}_T(x,0,t)] (-i) \Delta_2 \nonumber \\
&=& \frac{e^2}{(2\pi)^3} m (1-x) \Bigg[ (1-x)
(-\Delta_1-i \Delta_2) I_3+2 I_4 \Bigg]\,, \\
F^{1}_T(\downarrow\downarrow) & = & \frac{1}{2m}[E_T(x,0,t)+2
\tilde{H}_T(x,0,t)] (-i) \Delta_2 \nonumber \\
&=& \frac{e^2}{(2\pi)^3} m (1-x) \Bigg[ (1-x)
(\Delta_1-i \Delta_2) I_3-2 I_4 \Bigg]\,.
\end{eqnarray}
Here, $\uparrow \downarrow (\downarrow \uparrow) $ denotes the helicity flip of the 
electron and $\uparrow \uparrow (\downarrow \downarrow) $ means that initial state
has the same helicity as the final state. Only three of the above equations
are needed to disentangle the three unknowns, $H_T, E_T$ and $\tilde{H}_T$.
Similarly, the four different cases corresponding
to $j=2$ are given by,
\begin{eqnarray}
F^{2}_T(\uparrow\downarrow) & = &  H_T(x,0,t)
+\frac{\tilde{H}_T(x,0,t)}{2 m^2} (\Delta_1)(\Delta_1-i\Delta_2)
\nonumber \\
&=& \frac{e^2}{(2\pi)^3}\frac{x}{1-x}\left[I_1+I_2+C I_3\right]\,,
\\
F^{2}_T(\downarrow\uparrow) & = & -H_T(x,0,t)
+\frac{\tilde{H}_T(x,0,t)}{2 m^2} (-\Delta_1)(\Delta_1+i\Delta_2)
\nonumber \\
&=& -\frac{e^2}{(2\pi)^3}\frac{x}{1-x}\left[I_1+I_2+C
I_3\right]\,,
\\
F^{2}_T(\uparrow\uparrow) & = & \frac{1}{2m}[E_T(x,0,t)+2
\tilde{H}_T(x,0,t)](i \Delta_1) \nonumber \\
&=& \frac{e^2}{(2\pi)^3} m (1-x) \Bigg[ (1-x)
(i\Delta_1-\Delta_2) I_3+2 I_5 \Bigg]\,, \\
F^{2}_T(\downarrow\downarrow) & = & \frac{1}{2m}[E_T(x,0,t)+2
\tilde{H}_T(x,0,t)](i \Delta_1) \nonumber \\
&=& \frac{e^2}{(2\pi)^3} m (1-x)\Bigg[ (1-x)
(i \Delta_1+\Delta_2) I_3-2 I_5 \Bigg]\,,
\end{eqnarray}
where
\be
I_1 &=& \int {d^2 k_\perp \over L_1} = \pi log{\Lambda^2\over
m^2(1-x)^2}\,,\nonumber\\
I_2 &=& \int {d^2 k_\perp \over L_2}=\pi log{\Lambda^2\over
(m^2+\Delta^2_{\perp})(1-x)^2}\,, \nonumber\\
I_3 &=& \int {d^2 k_\perp \over L_1 L_2}= \pi \int_0^1 {d
\alpha\over D}\,, \nonumber \\
I_4 &=& \int {d^2 k_\perp (k_1)\over L_1 L_2}= \pi (1-x)\int_0^1
{(1-\alpha) \Delta_1 \over D} d \alpha\,, \nonumber \\
I_5 &=& \int {d^2 k_\perp (k_2)\over L_1 L_2}= \pi (1-x)\int_0^1
{(1-\alpha) \Delta_2 \over D} d \alpha\,;
\ee
and
\be
D &=&\alpha(1-\alpha)(1-x)^2 \Delta^2_{\perp}+ m^2 (1-x)^2\,
\nonumber \\
C &=& -2 m^2 (1-x)^2-(1-x)^2 \Delta_\perp^2\,, 
\ee
\be
L_1 = (k^\perp)^2-m^2 (1-x)^2, ~~~~L_2 = (k'^\perp)^2-m^2
(1-x)^2\,. 
\ee 
$\Lambda$ is the upper cutoff on transverse momentum. In \cite{marc}, a
lower cutoff, $\mu$ has been imposed on the transverse momentum, 
due to which the
logarithms  in $I_1$ and $I_2$ are of the form $log{\Lambda^2\over \mu^2}$.
As here we have imposed a cutoff on $x$ at $x \to 1$ instead, the cutoff on
$k_\perp$ is not necessary. $F_T^i( \uparrow \downarrow)$ and $F_T^i(
\downarrow \uparrow)$
receive contribution from the single particle sector of the Fock space,
which is of the form ${\cal N }^2\delta(1-x)$ where ${\cal N}^2$ is the 
normalization 
given by ${\mid \psi_1 \mid }^2$ in Eq. (3.6) of \cite{dip2}. As we exclude
$x=1$ by imposing a cutoff, we do not consider this contribution 
in this work.  However, 
the single particle contribution cancels the singularity
as $x \rightarrow 1$. This has been shown explicity in the forward limit in
\cite{tran}, namely for the transversity distribution $h_1(x)$. The
coefficient of the logarithmic term in the expression of $h_1(x)$ gives the
correct splitting function for leading order evolution of $h_1(x)$; the delta
function providing the necessary 'plus' prescription. In the off forward
case, the cancellation occurs similarly, as shown for $F(x, \xi,t)$ in
\cite{marc}. The behavior at $x=0,1$ can be improved by
differentiating the LFWFs with respect to $M^2$ \cite{har}.

The first moment of $(2 \tilde
H_T(x,0,0)+E_T(x,0,0))$ is normalized by
\be
\int_{-1}^{1} dx (2 \tilde H_T(x,0,0)+E_T(x,0,0))=\kappa_T
\ee
where $\kappa_T$ gives by how far the average position of
quarks with spin in ${\hat x}$ is shifted in ${\hat y}$ direction in an
unpolarized target relative to the transverse center of momentum. A sum rule
equivalent to Ji's sum rule has been derived in \cite{burchi}, for the
angular momentum $J^i$ carried by the quarks with transverse spin in an
unpolarized target. This is related to the second moment of the
chiral-odd GPDs, namely, $\int dx x [ H_T(x,0,0)+2 \tilde
H_T(x,0,0)+E_T(x,0,0)]$. $H_T(x,0,0)=h_1(x)$, the transversity distribution
in the forward limit. In addition to the overall shift in the transversity
asymmetry coming from $H_T(x,0,0)$, the other term containing $(2
\tilde H_T+E_T)$ gives the deformation in the center-of-momentum frame due
to spin-orbit correlation.

The transverse distortion in the impact parameter space given by the GPD $E$
has been shown to be connected with Sivers effect \cite{sivers} in a model
calculation \cite{burk1}. Similar connection with $\kappa_T$ and Boer-Mulders
effect \cite{mulders} has been suggested in \cite{burchi,chiral}. However, in
\cite{metz}, it has been shown that connection between transverse momentum
dependent parton distributions and ipdpdfs does not exist in a model
independent way.

As shown in \cite{chiral}, the term $(H_T- {t\over 4 m^2} \tilde H_T)$
represents the correlation between the transverse quark spin and the
transverse spin of the nucleon itself. Namely, the density of transversely
polarized quarks in a transversely polarized nucleon contain a term
proportional to $H_T-{t\over 4 m^2} \tilde
H_T $. In the forward limit $t=0$ and this reduces to the transversity
distribution $H_T(x,0,0)=h_1(x)$.

\begin{figure}[t]
\includegraphics[width=8cm,height=7cm,clip]{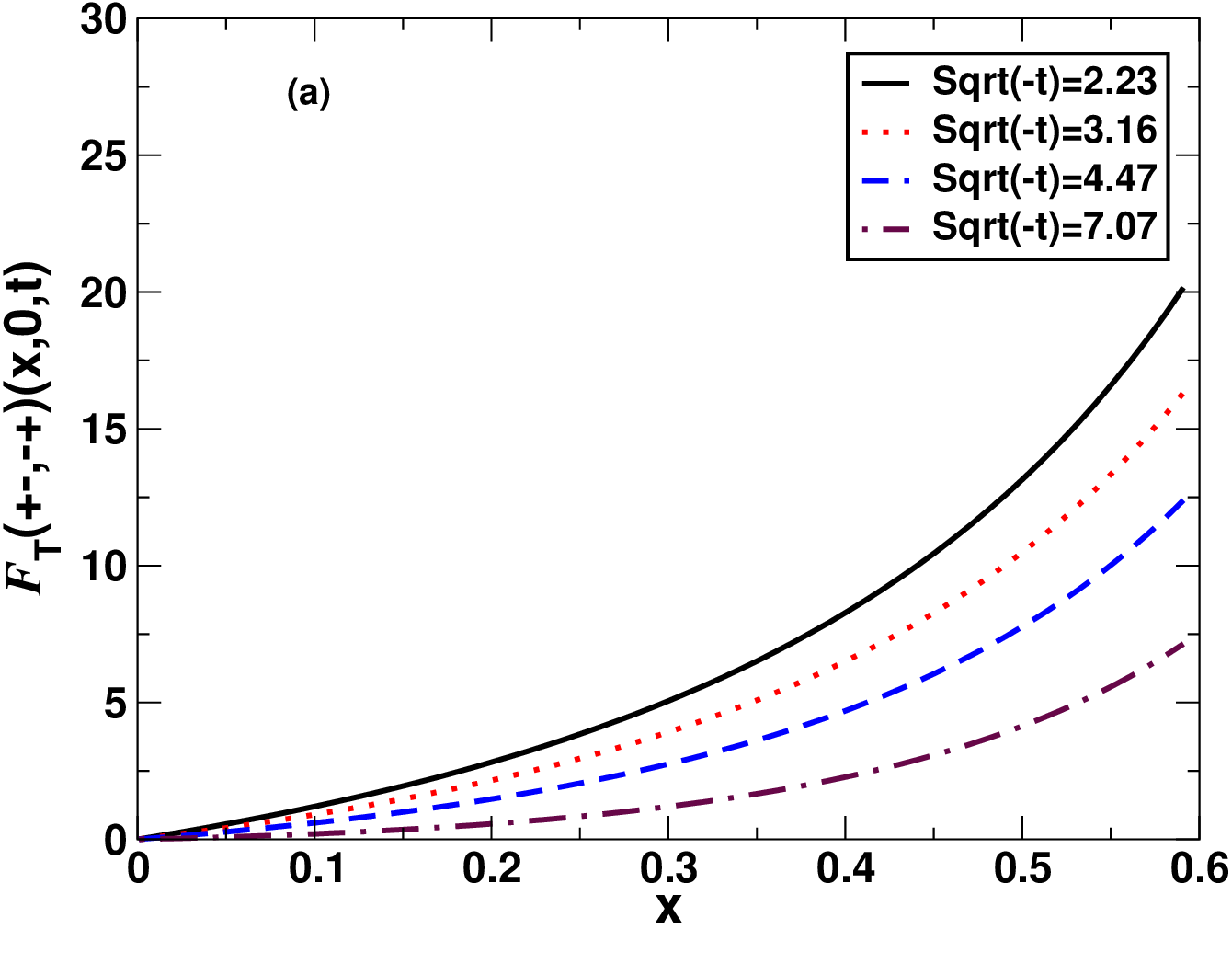}%
\hspace{0.2cm}%
\includegraphics[width=8cm,height=7cm,clip]{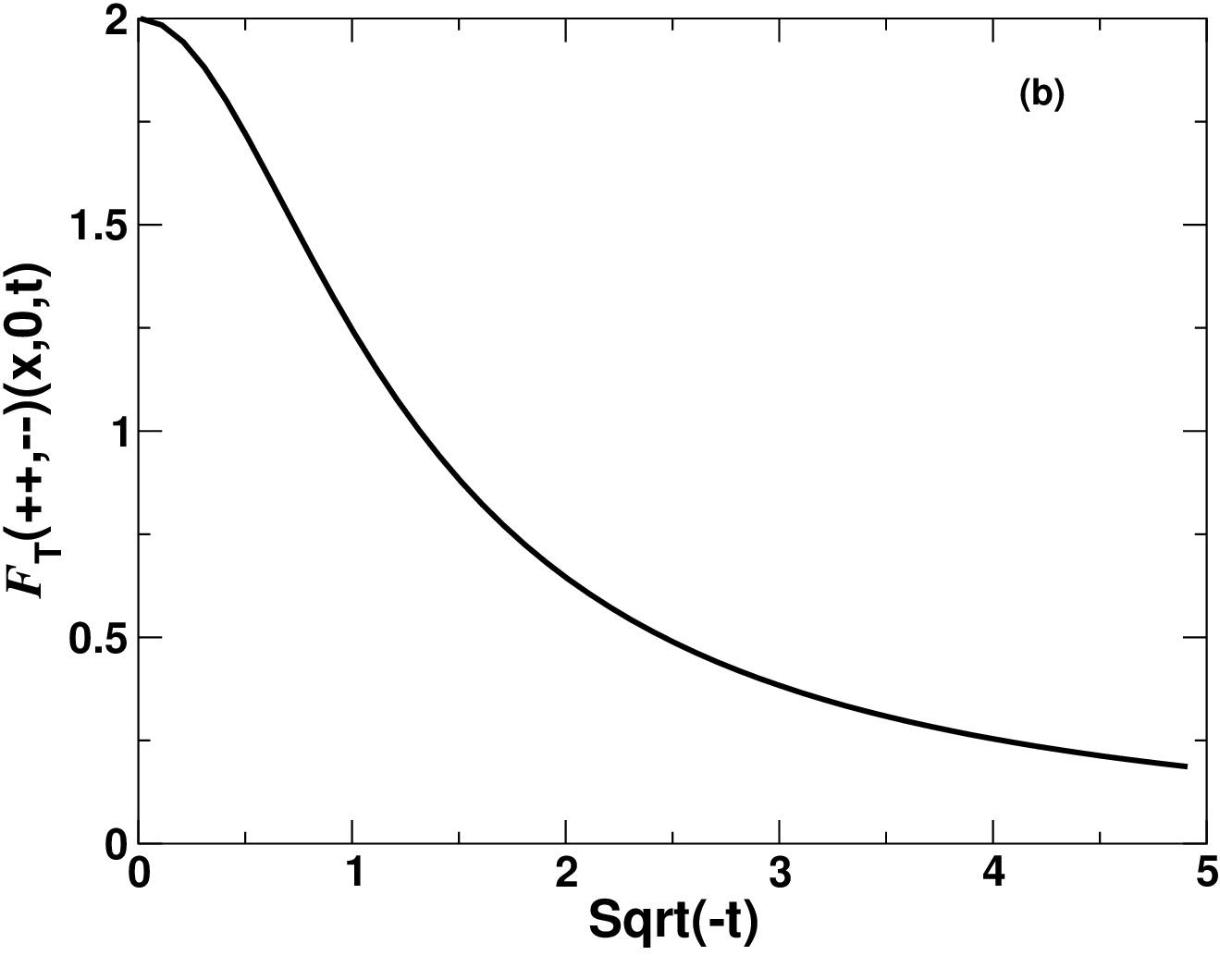}
\caption{Plots of  (a) $ F_T(+-,-+)(x,0,t)=\left[ H_T(x,0,t) +
\frac{\sum_i \Delta_i^2}{4m^2}\tilde{H}_T(x,0,t) \right]$ as a function of
$x$ for fixed values of $\sqrt{-t}$ in MeV and (b)
$F_T(++,--)(x,0,t)=\left [ E_T(x,0,t)+2 \tilde{H}_T(x,0,t)
\right]$ as a function of $\sqrt{-t}$ in MeV.}
\end{figure}

In order to extract different combinations of the chiral-odd GPDs
of phenomenological importance, we combine the above results to get,
\be
{\cal F}^1_T(+-,-+)(x,0,t)&=&\frac{F^{1}_T(\uparrow\downarrow)
+F^{1}_T(\downarrow\uparrow)}{2} = \left[ H_T(x,0,t) +
\frac{\Delta_2^2}{2m^2}\tilde{H}_T(x,0,t) \right] \nonumber
\\ &=& \frac{e^2}{(2\pi)^3}\frac{x}{1-x}  \left[
I_1+I_2+C I_3 \right]\, \label{htb1}
\ee
\be
{\cal F}^2_T(+-,-+)(x,0,t)&=&\frac{F^{2}_T(\uparrow\downarrow)
-F^{2}_T(\downarrow\uparrow)}{2} = \left[ H_T(x,0,t) +
\frac{\Delta_1^2}{2m^2}\tilde{H}_T(x,0,t) \right] \nonumber
\\ &=& \frac{e^2}{(2\pi)^3}\frac{x}{1-x}  \left[I_1+I_2+C I_3
\right]\, \label{htb2}
\ee
\be
{\cal F}^1_T(++,--)(x,0,t)&=&
\frac{F^{1}_T(\uparrow\uparrow)+F^{1}_T(\downarrow\downarrow)}{2}
= \left [ E_T(x,0,t)+2 \tilde{H}_T(x,0,t) \right ]{-i\Delta_2\over 2 m}
\nonumber
\\ &=& \frac{e^2}{(2\pi)^3}  m (1-x)^2 (-i) \Delta_2 I_3\,. \label{etb1}
\ee
\be
{\cal F}^2_T(++,--)(x,0,t)&=&
\frac{F^{2}_T(\uparrow\uparrow)+F^{2}_T(\downarrow\downarrow)}{2}
= \left [ E_T(x,0,t)+2 \tilde{H}_T(x,0,t) \right ]{i\Delta_1\over 2 m}
\nonumber
\\ &=& \frac{e^2}{(2\pi)^3}  m  (1-x)^2  (i\Delta_1)
I_3\,. \label{etb2} 
\ee 
The above equations show that $\tilde{H}_T(x,0,t)=0$ in this model. Analytic 
expressions for
$H_T(x,0,t), E_T(x,0,t)$ and $\tilde{H}_T(x,0,t)$ are given in the appendix
of \cite{metz} in a quark model in terms of $k_T$ integrals, where $k_T$ is
the transverse momentum of the quark. Apart from the overall color factors,
our results agree with them.  

In fig. 1 (a), we have plotted
$H_T(x,0,t) + \frac{\sum_i \Delta_i^2}{4m^2}\tilde{H}_T(x,0,t)
$ for fixed values of $t=-\Delta_\perp^2$ and as a function of $x$. 
It increases with x at fixed $t$, the magnitude
decreases with increasing $\Delta_\perp^2$. We have taken
${e^2\over (2 \pi)^3}=1$ and $m= 0.5$~MeV. The helicity flip
contributions depend on the scale $\Lambda$ which we have taken as
$100$~ MeV. This is similar to the unpolarized distribution
\cite{dip,dip2}. Note that the linear combination of the light-front
helicity states here gives the overlap matrix element for the
transversely polarized state. It is to be noted that both Eqs.
(\ref{htb1}) and (\ref{htb2}) reduce to the transversity
distribution $h_1(x)$ in the forward limit calculated in
\cite{tran}. Fig. 1(b) shows the plot of  $E_T(x,0,t)+2 \tilde{H}_T(x,0,t)$
as a function of $\sqrt{-t}$ in MeV. Note that
this quantity becomes $x$ independent because of the $(1-x)^2$ present in the
denominator of $I_3$. This is a particular feature of the model considered. 
\section{Impact space representation}

\begin{figure}[t]
\includegraphics[width=8cm,height=7cm,clip]{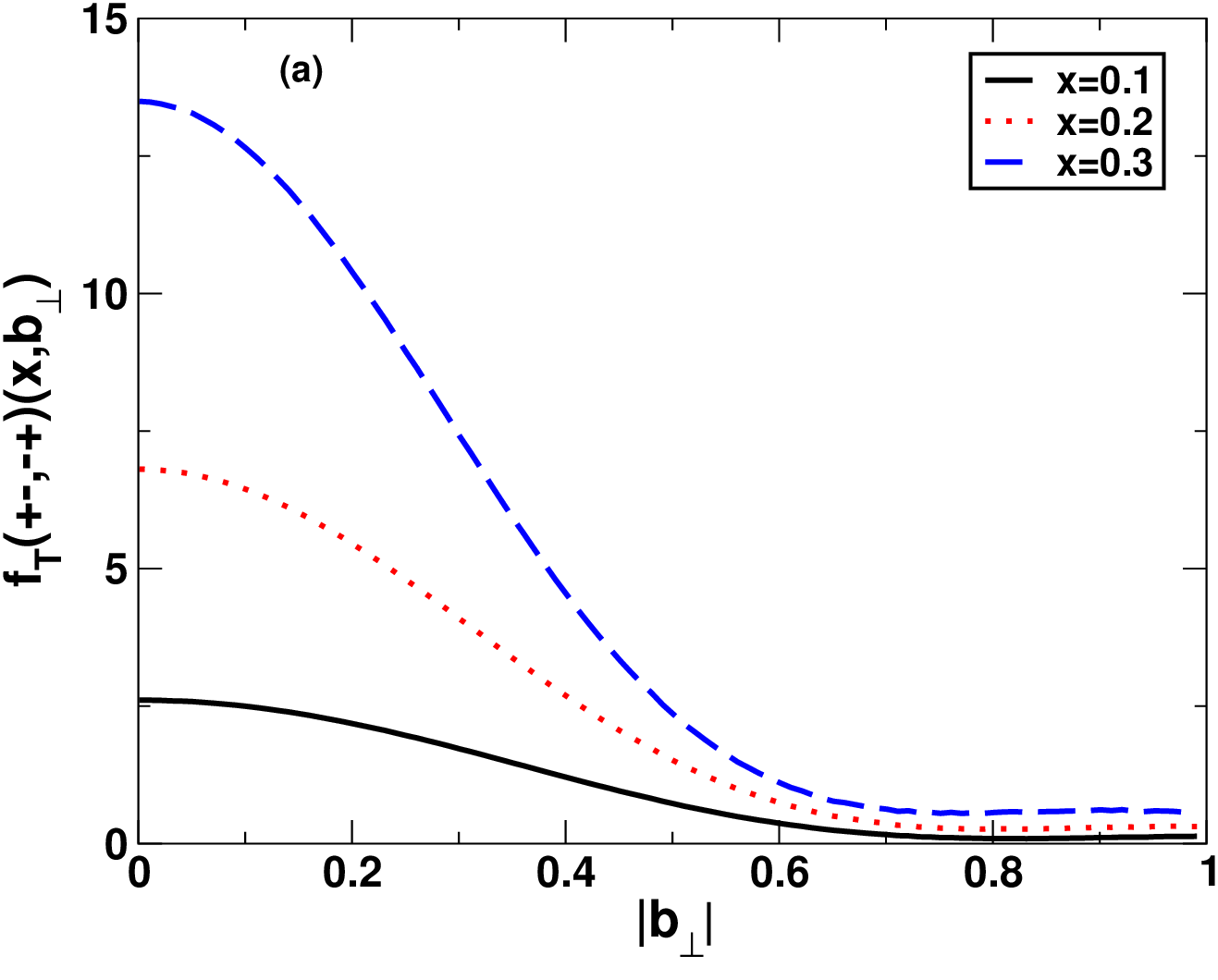}%
\hspace{0.2cm}%
\includegraphics[width=8cm,height=7cm,clip]{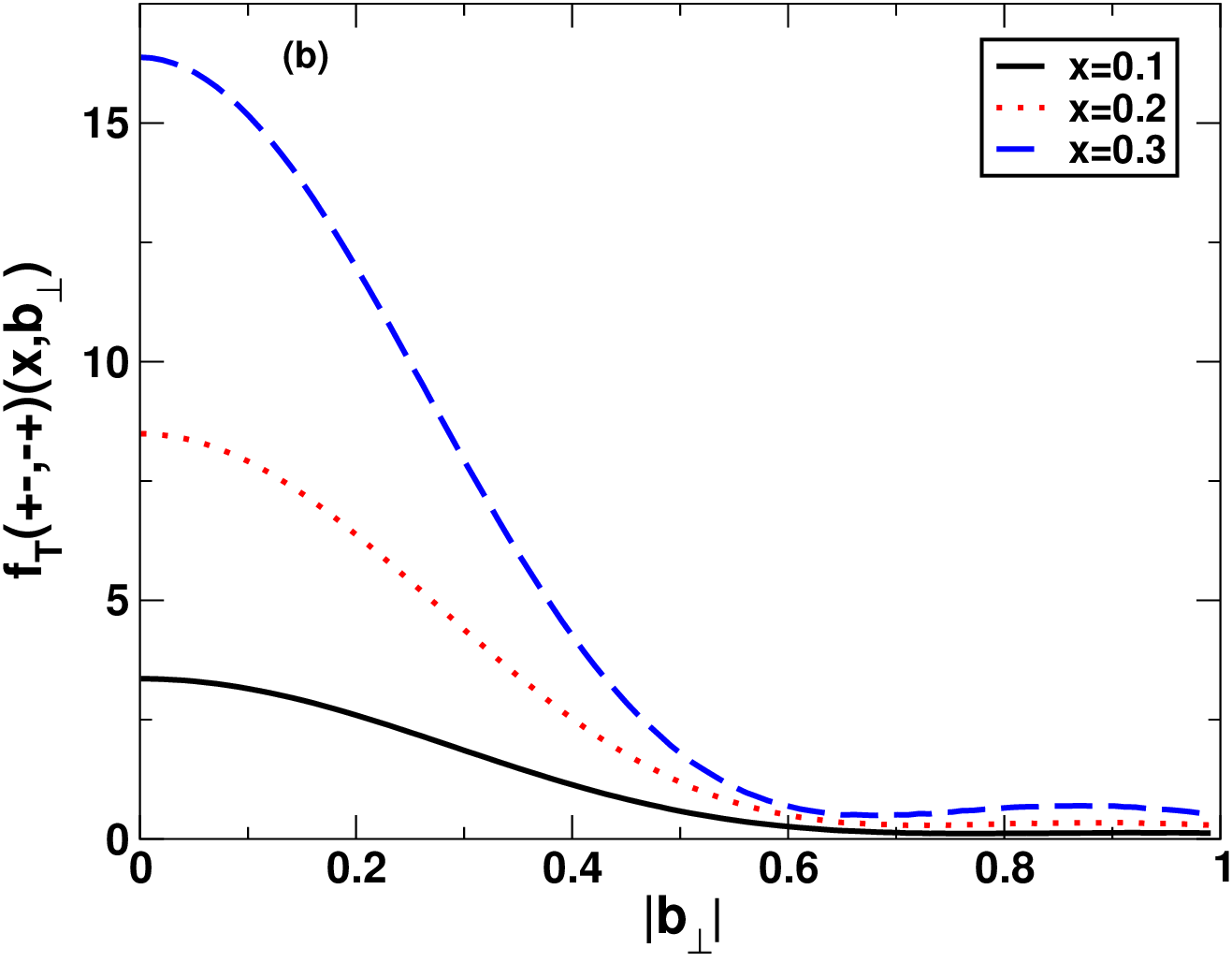}%
\caption{Impact parameter dependent pdfs $f_T(+-,-+)(x,b_\perp)=
\left[
   \mathcal{H}_T(x,b_\perp) -
    \frac{\Delta_b}{4m^2}\tilde \mathcal{H}_T(x,b_\perp) \right]$ as a
  function of $\mid b_\perp \mid $ for fixed
values of $x$ at (a) $m=0.5$~MeV and (b) $m=0.8$~MeV. $\mid b_\perp \mid$ is
in $\mathrm{MeV}^{-1}$. }
\end{figure}

The impact parameter dependent parton distributions are defined
from the GPDs by taking a Fourier Transform (FT) in $\Delta_\perp$ as follows :
\be
\mathcal{H}_T(x,b_\perp)&=& {1\over (2 \pi)^2} \int d^2 \Delta_\perp e^{-i
b_\perp \cdot \Delta_\perp} H_T(x,0, -\Delta_\perp^2)\,,\nonumber\\
\ee
\be
\mathcal{E}_T(x,b_\perp)&=& {1\over (2 \pi)^2} \int d^2 \Delta_\perp e^{-i
b_\perp \cdot \Delta_\perp} E_T(x,0, -\Delta_\perp^2)\,,\nonumber\\
\ee
\be
\tilde \mathcal{H}_T(x,b_\perp)&=& {1\over (2 \pi)^2} \int d^2 \Delta_\perp 
e^{-i
b_\perp \cdot \Delta_\perp} \tilde H_T(x,0, -\Delta_\perp^2)\,,\nonumber\\
\ee
where $b_\perp$ is the impact parameter conjugate to
$\Delta_\perp$.

We can write,
\be
f_T(+-,-+)(x, b_\perp)&=&\int {d^2 \Delta_\perp\over (2 \pi)^2}
e^{-i b_\perp. \Delta_\perp} \left[ H_T (x,0, -\Delta_\perp^2) +
\frac{\Delta^2_i}{4m^2}\tilde{H}_T (x,0, -\Delta_\perp^2) \right]
\nonumber\\&=& \left[ \mathcal{H}_T(x,b_\perp) -
\frac{\Delta_b}{4m^2}\tilde \mathcal{H}_T(x,b_\perp)
\right]\nonumber\\ &=&\frac{e^2}{2(2\pi)^3}\frac{x}{1-x}
\int_0^\infty \Delta d \Delta
   \left[
log{\Lambda^2\over m^2(1-x)^2}\right. \nonumber \\&& \left.+
log{\Lambda^2\over(m^2+\Delta^2_{\perp})(1-x)^2}+C \int_0^1 {d
  \alpha\over D} \right] J_0(b \Delta) \,.
\ee
where
\be
\Delta_b f={\partial \over
\partial b_i}{\partial \over
\partial b_i}f\,.
\ee 
$J_0(b \Delta)$ is the Bessel function. 
In fig. 2 we have plotted $f_T(+-,-+)(x,b_\perp)$ as a
function of $b_\perp$ for fixed $x$ for two different values of
the mass parameter $m$. It is peaked at $b_\perp=0$ and falls away
further from it. The peak increases as $x$ increases. This
quantity describes the correlation between the transverse quark
spin and the target spin in a transversely polarized nucleon in
impact parameter space. For an elementary Dirac particle, this
would be a delta function at $b_\perp=0$. The smearing in
transverse position space occurs due to the two-particle LFWF.

\begin{figure}[t]
\includegraphics[width=8cm,height=7cm,clip]{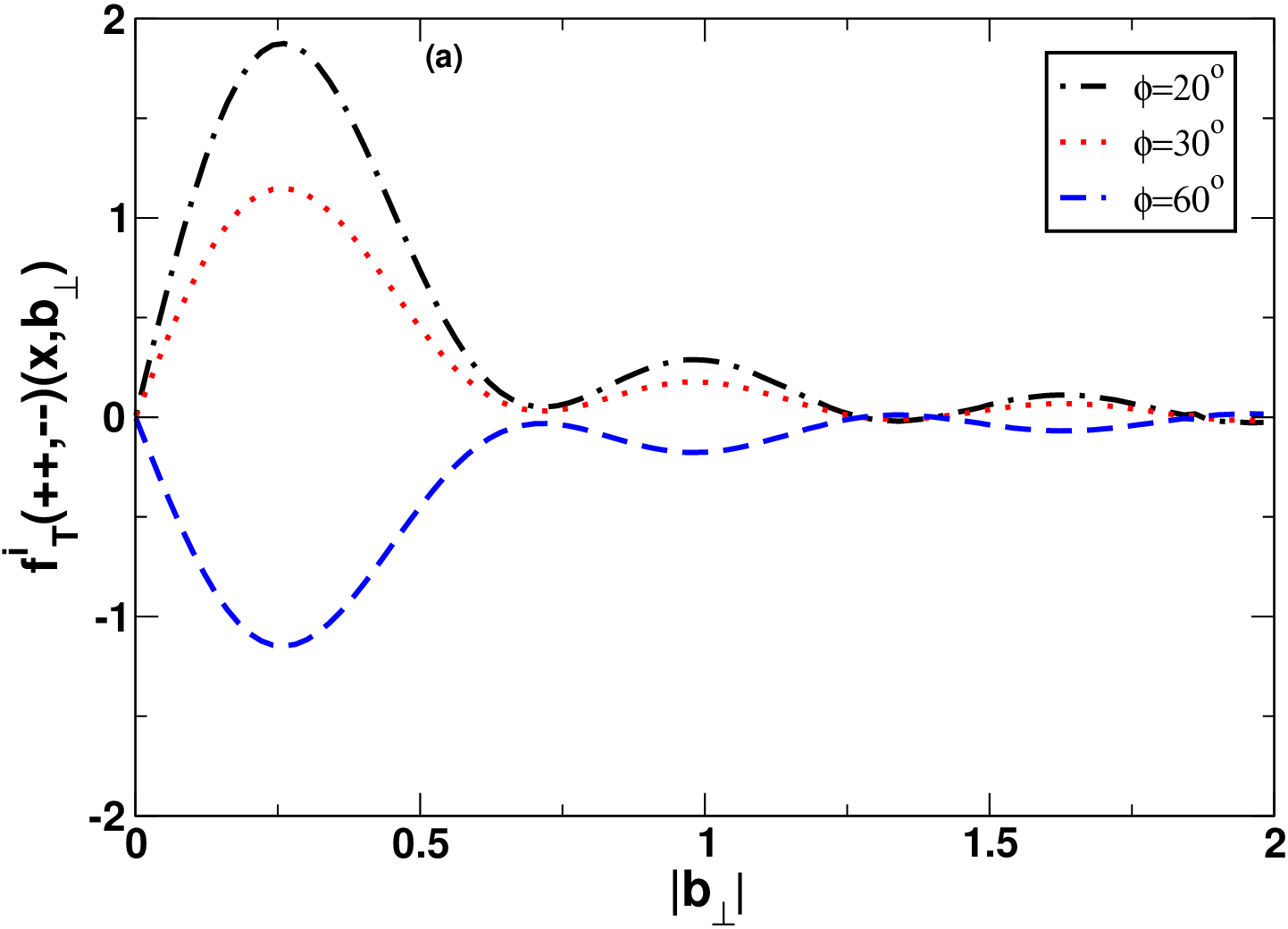}
\hspace{0.2cm}%
\includegraphics[width=8cm,height=7cm,clip]{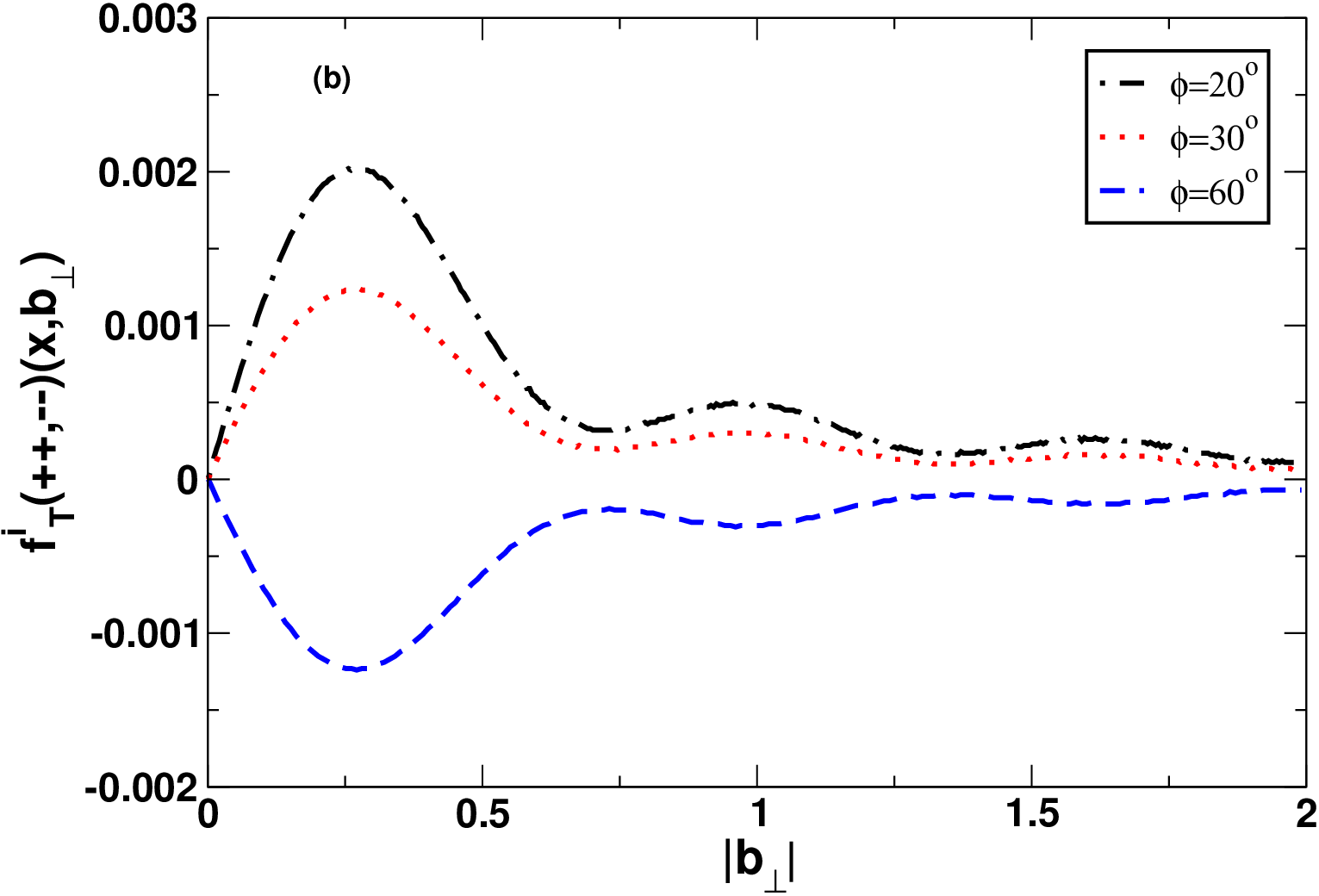}%
\caption{Impact parameter dependent pdfs $f^i_T(++,--)(x,b_\perp)
=-\epsilon^{ij}b_j{\partial \over
\partial B}\left [ \mathcal{E}_T(x,b_\perp)+2 \tilde {\mathcal{H}}_T
(x,b_\perp) \right ]$ as a function of $\mid b_\perp \mid $ for 
different values of $\phi$ and (a) $m=0.5$~MeV and (b) $m=0.01$~MeV
. $\mid b_\perp \mid$ is in $\mathrm{MeV}^{-1}$.}
\end{figure}

The transversity density of quarks is defined as
\be
\delta^i q(x,b_\perp)= -{\epsilon^{ij}\over m} b_j{\partial \over \partial
b^2} ( 2 \tilde \mathcal{H}_T +\mathcal{E}_T)
\ee
Even for an unpolarized target, it can be non-zero (as shown in
\cite{burchi} using in a simple model). This is due to spin-orbit
correlations in the quark wave function. If the quarks have orbital angular
momentum then their distribution is shifted to one side. In an unpolarized
nucleon, all orientations are equally probable and therefore, the
unpolarized distribution is axially symmetric. However, if there is a
spin-orbit correlation, then quarks with a certain spin orientation will
shift to one side and those with a different orientation will shift to
another side.

This can be constructed from,
\be
f^i_T(++,--)(x,b_\perp)&=&-\epsilon^{ij}b_j{\partial \over
\partial B}\left [ \mathcal{E}_T(x,b_\perp)+2 \tilde \mathcal{H}_T
(x,b_\perp) \right ] \nonumber \\ & = & i\epsilon^{ij}\int {d^2
\Delta_\perp\over (2 \pi)^2} \Delta_j e^{-i b_\perp. \Delta_\perp}
\left [ E_T(x,0, -\Delta_\perp^2)+2 \tilde{H}_T (x,0,
-\Delta_\perp^2) \right ],\nonumber\\ &=& -i{\epsilon^{ij}b_j\over
b}\int {(\Delta)^2\, d \Delta\over (2 \pi)} \left [ E_T(x,0,
-\Delta_\perp^2)+2 \tilde{H}_T (x,0, -\Delta_\perp^2) \right ]
J_1(b \Delta )\,, 
\ee 
where ${\partial \over \partial B}= 2{\partial \over \partial b^2}$ and   
\be
b_1=b \cos \phi\,,~~~~~~b_2=b \sin \phi\,.
\ee
For computational purpose, we have used
\be
J_n(b \Delta )={1\over \pi}\int_0^\pi d\theta \cos(n \theta-b
\Delta \sin\theta).
\ee
In Fig. 3, we have plotted $f^i_T(++,--)(x,b_\perp)$ as a function of
$b_\perp$ for different values of $\phi$. As stated before, in the simple
model we consider, this quantity is independent of $x$. We took
the constant phase factor $(-i)$ out. The
effect of this term is to shift the peak of the impact parameter space 
density (see eq. (8) of \cite{chiral}) away from
$b_\perp=0$. This shift clearly shows the interplay between the spin and the
orbital angular momentum of the constituents of the two-particle LFWF.  
In order to understand the plots, consider a two-dimensional plane with 
$b_1$ and $b_2$ plotted along the two
axes. Fixed $\mid b_\perp \mid$ denote concentric circles in this plane.
From our plots, we see that the position of the peak of 
$ f_T^i $ is indepenent of $\phi$ and $m$. The
magnitude of the peak increases as $m$ increases.  
The magnitude and sign changes as $\phi$ changes. 
$f_T^i $ would vanish at $\phi= {\pi\over 4}$ and ${5 \pi\over 4}$. 
In 2-D $b_1-b_2$ plane, the primary peak will lie on a circle and the
secondary peak will lie on a concentric circle with larger radius.
\section{conclusion}
In this work, we have studied the chiral-odd GPDs in impact parameter space
in a self consistent relativistic two-body model, namely for the quantum fluctuation
of an electron at one loop in QED. In its most general form \cite{drell}, 
this model can act as a template for the
quark-spin one diquark light front wave function for the proton, although
not numerically. Working in
light-front gauge, we expressed
the GPDs as overlaps of the light-front wave functions. We took the skewness
to be zero. Only the diagonal $2 \to 2 $ overlap contributes in this case.
The impact space representations are obtained by taking Fourier transform of
the GPDs with respect to the transverse momentum transfer. It is known
\cite{chiral,burchi} that certain combinations of the chiral-odd
GPDs in impact parameter space affect the quark and nucleon spin
correlations in different ways. For
example, the combination $\mathcal{H}_T- {\Delta_b\over 4 m^2} \tilde
\mathcal{H}_T$ gives the correlation between the transverse quark spin and the
target spin in a transversely polarized nucleon. On the other hand,
the quantity $ \epsilon_{ij} b_j {\partial \over \partial B}( \mathcal{E}_T+
2 \tilde \mathcal{H}_T)$ gives the spin-orbit correlation of the quarks in
the nucleon. We have investigated both and have shown that due to the
interplay between the spin and orbital angular momentum of the 2-particle
LFWF, the distribution in the impact parameter space is shifted sideways.
In a future work, we plan to investigate the various positivity constraints
for the chiral-odd GPDs as well as the effect of non-zero skewness $\xi$,
when there is a finite momentum transfer in the longitudinal direction as
well.
\section{acknowledgment}
AM thanks DST Fasttrack scheme, Govt. of India  for financial support
for completing this work.


\begin{thebibliography}{99}

\bibitem{rev} For reviews on generalized parton distributions,
and DVCS, see M. Diehl,
Phys. Rept, {\bf 388}, 41 (2003); A. V. Belitsky and A. V. Radyushkin, Phys.
Rept. {\bf 418} 1, (2005);  K. Goeke, M.
V. Polyakov, M. Vanderhaeghen, Prog. Part. Nucl. Phys. {\bf 47}, 401 (2001).

\bibitem{bur} M. Burkardt, Int. Jour. Mod. Phys. {\bf A 18}, 173 (2003);
M. Burkardt, Phys. Rev. {\bf D 62}, 071503 (2000), Erratum-
ibid, {\bf D 66}, 119903 (2002); J. P. Ralston and B. Pire, Phys. Rev. {\bf
D 66}, 111501 (2002).

\bibitem{soper} D. E. Soper, Phys. Rev. D {\bf 15}, 1141 (1977).



\bibitem{burchi} M. Burkardt, Phys. Rev {\bf D 72}, 094021 (2005).

\bibitem{ji} X. Ji. Phys. Rev. Lett. {\bf 78}, 610 (1997).

\bibitem{markus} M. Diehl, Eur. Phys. J. C {\bf 19}, 485 (2001).

\bibitem{chiral} M. Diehl and P. Hagler, Eur.Phys.J.{\bf C 44}, 87
(2005).

\bibitem{double} D. Yu Ivanov, B. Pire, L. Szymanowski, O. V. Teryaev, Phys.
Lett. {\bf B 550}, 65 (2002); Phys. Part. Nucl. {\bf 35}, 67
(2004).

\bibitem{barbara} B. Pasquini, M. Pincetti, S. Boffi, Phys. Rev. {\bf D 72},
094029 (2005).

\bibitem{drell} S. J. Brodsky and S. D. Drell, Phys. Rev. {\bf D 22}, 2236
(1980).

\bibitem{dis} A. Harindranath, R. Kundu, W. M. Zhang, Phys.
Rev. {\bf D 59}, 094013 (1999);  A. Harindranath, A. Mukherjee,
R. Ratabole, Phys. Lett. {\bf B 476},  471 (2000); Phys. Rev.
{\bf D 63}, 045006 (2001).

\bibitem{dip} D. Chakrabarti and A. Mukherjee, Phys. Rev. {\bf D 71}, 014038
(2005). 

\bibitem{dip2} D. Chakrabarti, A. Mukherjee, Phys. Rev. {\bf D 72}, 
034013 (2005).

\bibitem{marc} A. Mukherjee and M. Vanderhaeghen, Phys. Lett. {\bf B 542}, 245 (2002);
Phys. Rev. {\bf D 67}, 085020 (2003).

\bibitem{si} S. J. Brodsky, D. Chakrabarti, A. Harindranath, A.
Mukherjee, J. P. Vary, Phys. Lett. {\bf B, 641}, 440 (2006), Phys. Rev. {\bf
D 75}, 0143003 (2007).

\bibitem{orbit} S. J. Brodsky, D. S. Hwang, B-Q. Ma, I. Schmidt, Nucl. Phys.
{\bf B 593}, 311 (2001).

\bibitem{overlap}  S. J. Brodsky, M. Diehl, D. S. Hwang, Nucl. Phys. {\bf B
596}, 99 (2001); M. Diehl, T. Feldmann, R. Jacob, P. Kroll, Nucl. Phys.
{\bf B 596}, 33 (2001), Erratum-ibid {\bf  605}, 647 (2001).

\bibitem{two} W. M. Zhang, A. Harindranath, Phys. Rev. {\bf D 48}, 4881
(1993).


\bibitem{tran} A. Mukherjee and D. Chakrabarti,
Phys. Lett. {\bf B 506}, 283 (2001).

\bibitem{har} H. Dahiya, A. Mukherjee, S. Ray, Phys. Rev. {\bf D 76}, 034010
(2007). 


\bibitem{sivers} D. W. Sivers, Phys. Rev. {\bf D 43}, 261 (1991).

\bibitem{burk1} M. Burkardt, Phys. Rev. {\bf D 66}, 114005 (2002).

\bibitem{mulders} D. Boer and P. J. Mulders, Phys. Rev. {\bf D 57}, 5780
(1998).

\bibitem{metz} S. Meissner, A. Metz and K. Goeke, hep-ph/0703176.



\end{thebibliography}
\end{document}